\documentclass[preprint2]{aastex}

\slugcomment{Submitted to PASP}
\shorttitle{Station-Keeping Requirements for Imaging Interferometers in Space}
\shortauthors{Ronald J.\ Allen}

\begin{document}


\title{Station-Keeping Requirements \\
for Constellations of Free-Flying Collectors \\
Used for Astronomical Imaging in Space\\[0.2in]
}

\author{Ronald J. Allen}
\affil{Space Telescope Science Institute, 3700 San Martin Drive, 
Baltimore, MD 21218}
\email{rjallen@stsci.edu}

\begin{abstract}
The accuracy requirements on station-keeping for constellations of free-flying
collectors coupled as (future) imaging arrays in space for astrophysics
applications are examined. The basic imaging element of these arrays is the
two-element interferometer. Accurate knowledge of two quantities is required:
the \textit{projected baseline length}, which is the distance between the two
interferometer elements projected on the plane tranverse to the line of sight
to the target; and the \textit{optical path difference}, which is the
difference  in the distances from that transverse plane to the beam combiner.
``Rules-of-thumb'' are determined for the typical accuracy
required on these parameters. The requirement on the projected baseline length 
is a \textit{knowledge} requirement and depends on the angular size of the
targets of interest; it is generally at a level of half a meter for typical
stellar targets, decreasing to perhaps a few centimeters only for the
widest attainable fields of view. The requirement on the optical path difference
is a \textit{control} requirement and is much tighter, depending on the
bandwidth of the  signal; it is at a level of half a wavelength for narrow
(few \%) signal bands, decreasing to $\approx 0.2 \, \lambda$ for the broadest
bandwidths expected to be useful. Translation of these requirements
into engineering requirements on station-keeping accuracy depends on the
specific details of the collector constellation geometry. Several examples
are provided to guide future application of the criteria presented here.
Some implications for the design of such collector constellations and for the
methods used to transform the information acquired into images are discussed.
\end{abstract}


\section{Introduction}

One of the major problems affecting the design of future systems for
high-resolution astronomical imaging using constellations of free-flying 
collectors in space\footnote{Some examples of such systems presently under
study by NASA include the ``Stellar Imager'' (SI), the ``Terrestrial Planet
Finder - Interferometer'' (TPF-I), and the SPECS mission study for a
sub-millimeter space interferometer.} is the necessity to precisely
maintain the ``optical figure'' or surface accuracy of the equivalent aperture
for extended periods of time. On sufficiently-bright targets, this
accuracy might be achieved using signal photons from the target itself in order
to operate various control loops. Such ``adaptive'' control systems can be found,
for instance, driving deformable mirrors on filled-aperture telescopes, and
controlling delay lines on interferometers. However, in the more general (and
often more interesting) case where the target is very faint, the accuracy
requirement translates into a tight requirement on the geometry of the
optical system. In the case of a constellation of free-flying collectors in
space, the accuracy requirement on the optical figure of the instrument for
observations of faint targets becomes a requirement on
\textit{station-keeping}.

Station-keeping may be either \textit{inertial}, with respect to some
fixed reference points (such as distant quasars), or \textit{relative} to the
rest of the spacecraft in the constellation. Relative station-keeping
may be done so as to
yield a good optical figure, but still leave the whole constellation tumbling
about some arbitrary axis, so this alone is not sufficient. A significant
design effort has been expended on dealing with the ``inertial'' part of the
problem, and most approaches for measuring the overall rotation of the whole constellation involve the use of the ``rotating Sagnac interferometer''
described e.g.\ by \citet{hec02}\footnote{One design using this concept is
called the ``Kilometric Optical Gyro''.}.

Early studies \citep{jon91, esa96} of the station-keeping problem with
free-flying collectors took as a ``straw-man'' criterion that the
positional accuracy required would
be of the order of a small fraction of a wavelength, e.g.\ $\approx \lambda/50$.
This translates to $\approx 10$ nm at an optical wavelength of 500 nm.
A combination of radio and laser-ranging systems may be adequate for
station-keeping at a level of microns, but more elaborate measures will be
required in order to reach the nanometer level if such accuracy is really
required. Current thoughts for achieving
this level of accuracy in the absence of sufficient photons from the
target itself include fine control using photons from another
nearby bright ``phase reference'' star in the field of view,
or the development of major technical improvements in ranging systems.

\begin{figure*}[ht!]
\epsscale{1.5}
\plotone{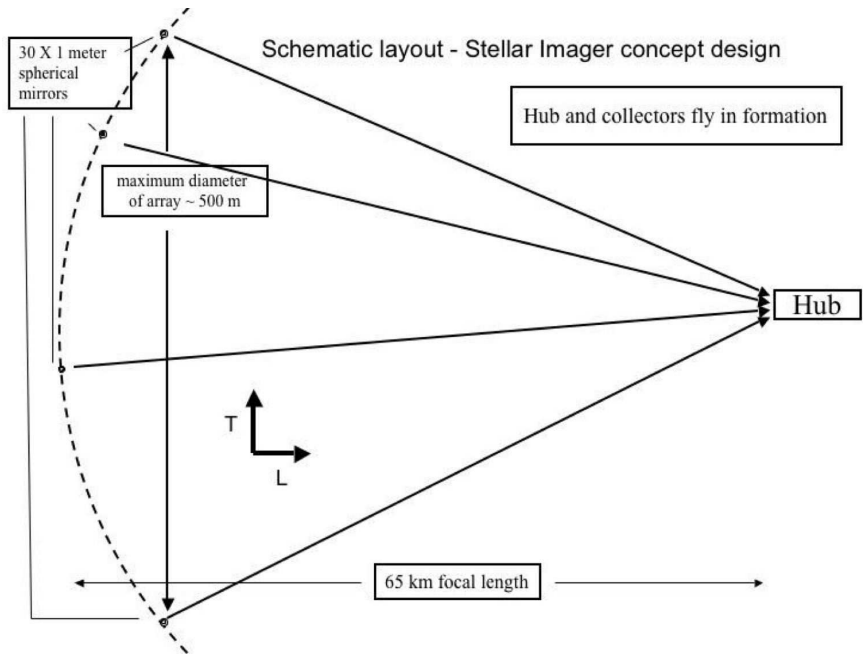}
\caption{
Side view sketch of a possible design for the ``Stellar Imager'' concept
(not to scale). The collectors, which are identical 1-meter spherical mirrors
in this design, reflect the light from the target (off to the far right in the
figure in the direction indicated by the arrow ``L'') to a central hub
spacecraft. \label{fig:SIa-sketch}}
\end{figure*}

The use of a phase reference star has been developed extensively in
order to overcome severe fluctuations in the fringe pattern caused by 
atmospheric turbulence in ground-based optical and near-IR long-baseline
interferometers (see e.g.\ the summary by \cite{qui00}). Application to the
(more liesurely but analogous) problem of station-keeping in a slowly-wandering
space interferometer constellation was proposed e.g.\ in an ESA study more than
ten years ago \citep{esa96} where it was shown that the tolerances in the
directions perpendicular and parallel to the direction to the target could be
reduced by factors of order $10^{3}$ and $10^{6}$, respectively, if a
bright phase-reference star existed within an angular distance of $2'$
from the target of interest. However, the penalty here is a reduction in that
fraction of the celestial sphere which can be observed, since suitable
phase-reference stars may not be available for every target of interest. In
addition, the effective field of view of the constellation's optical system
is likely to be very small\footnote{For example, this would be the case
with certain designs for the Stellar Imager (Figure \ref{fig:SIa-sketch}).},
thereby further reducing the utility of this approach.

Compared to the ground-based atmospheric-turbulence problem, the longer time
scales expected for station-keeping drifts may make it feasible to use photons
from the target itself to stabilize the fringes. This can be done in certain
specialized applications (e.g.\ nulling interferometers such as TPF-I and
Darwin), but becomes increasingly difficult in general for faint targets. In
addition, the target should not be appreciably resolved at the interferometer
baselines in use, since the signal photon level then also decreases. But the
main science goal may well be to measure just this kind of structure in the
target; hence the need to use target photons for fringe stabilization is in
general very restrictive.

Since we can expect that current laser-ranging technology will continue to
improve\footnote{Impressive gains in precision laser metrology have recently
been achieved for use in the NASA/JPL Space Interferometry Mission (SIM).},
one can ask whether it might be possible to relax the
$\approx \lambda/50$ requirement enough to accomplish station-keeping purely
by on-board laser-ranging methods. This would permit the observation of
arbitrarily-faint targets located anywhere in the sky. Questions which arise
include: How firm is the ``$\lambda/50$'' requirement? Does the physics of
producing images from interferometer fringes justify this number? Are there
ways of extracting the desired image information, perhaps only approximately,
but which can tolerate larger errors? The search for answers to these questions
has motivated the work described in this paper.

\begin{figure*}[ht!]
\epsscale{1.5}
\plotone{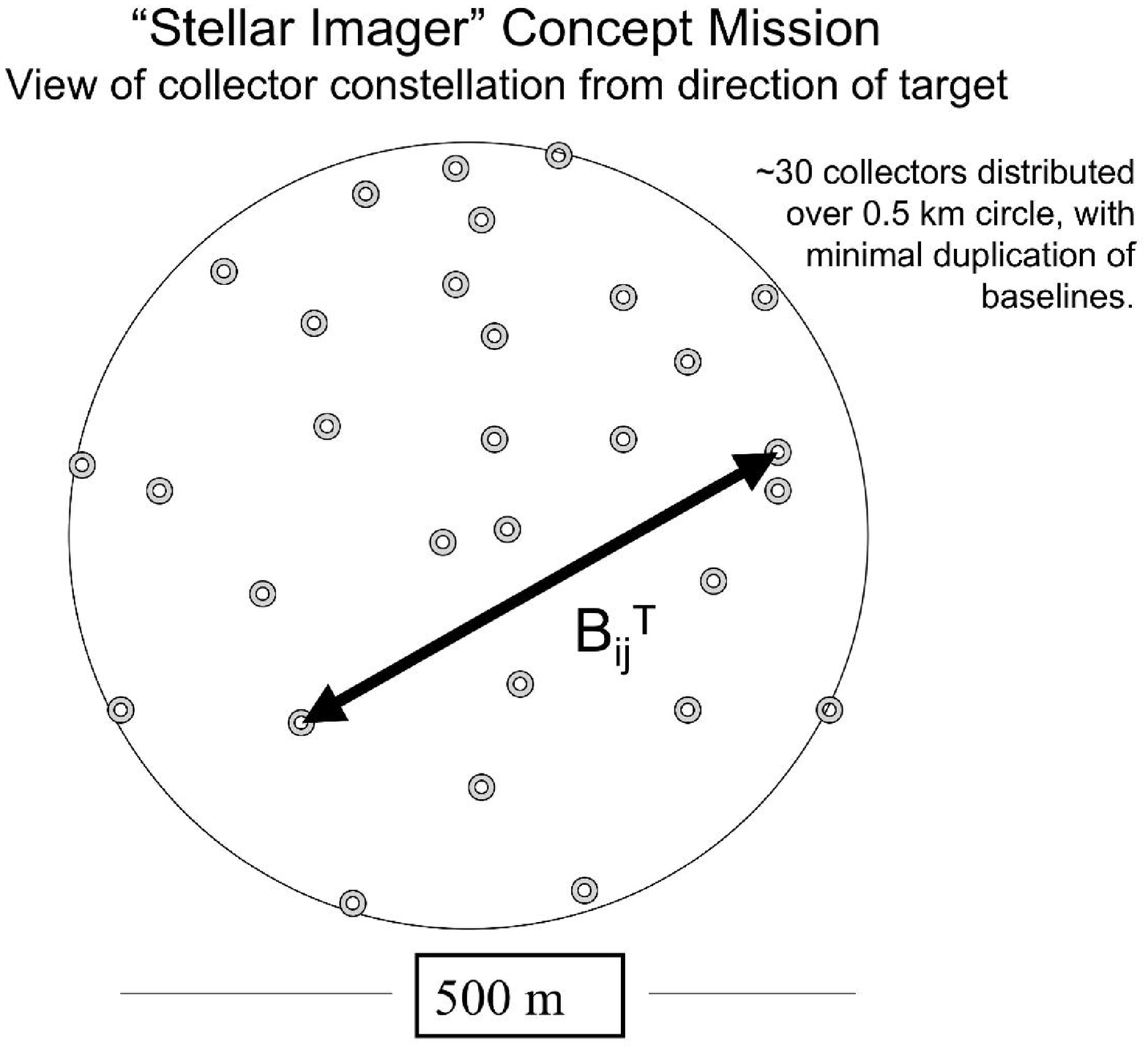}
\caption{
View of the ``Stellar Imager'' collector constellation from the target showing
one of the many interferometer baselines which make up this telescope
``aperture''. This view is in the transverse plane (indicated by the arrow ``T'' 
in Figure \ref{fig:SIa-sketch}), and B$_{ij}^{T}$ is the distance between any
two collectors projected on this plane. \label{fig:SIb-sketch}
}
\end{figure*}

An example of the kind of space-based imaging system considered here is shown in
Figure \ref{fig:SIa-sketch}. This is a cross-section sketch of a concept design
for the Stellar Imager, a constellation of about 30 free-flying collectors
operating at optical/UV wavelengths and distributed over an area with maximum
dimension of up to 500 meters\footnote{The Stellar Imager mission study is
described at: \\
http://hires.gsfc.nasa.gov/$\sim$si.}. These collectors direct the
radiation they receive to a central ``hub'' spacecraft for further processing
and downlink. Figure \ref{fig:SIb-sketch} shows a possible distribution of the
collectors as seen from the central hub.

The plan of this paper is as follows. First, a simple (but entirely general)
model of the imaging process for free-flying constellations of collectors is
described. The basic imaging element of these constellations is the two-element
interferometer, and the image quality is particularly sensitive to errors in the 
determination (and stability) of the phase of the interferometer fringe pattern. 
This model leads to the identification of two important parameters which play a
role in determining the accuracy with which the imaging can be done. The first 
of these is the \textit{projected baseline length}, which is the distance
between the two collector elements of each constituent interferometer projected
on a plane perpendicular to the line of sight to the target, and the second 
is the \textit{optical path difference}, which is the difference in
the distances from that perpendicular plane through each collector to the
interferometer beam combiner.

At this point it is useful to clarify the usage here of
the terms \textit{knowledge} of a parameter and its \textit{control}.
In general, precise control of baseline length is
not required; what \textit{is} required is precise knowledge of the actual
values at any time. That knowledge permits us to place a data sample at the
right location in the aperture plane before carrying out the imaging
computations. On the other hand, precise knowledge of the
optical path difference is not sufficient; we must control this quantity to
be close to zero at all times. Just how this knowledge and control will be
provided will depend on the design of the
particular constellation; those details actually do not concern us here, but
they will affect the distribution of the error budget over the various parts
of the instrument. What \textit{does} concern us here is to determine the
required accuracy on knowledge of the projected baseline length and
control of optical path difference in order to achieve a given level of
accuracy in the measurements which will go into the final synthesized image.
As we shall see, the physics of the interferometric imaging process implies
that the typical knowledge accuracy required for the projected baseline length
$B_{ij}^{T}$ turns out to depend on the angular size of the target of interest;
it is generally at a level of half a meter for typical stellar targets,
decreasing to a few centimeters only for the widest practical
fields of view. On the other hand, the control accuracy required for the
optical path difference OPD is much higher, and depends on the bandwidth of the
signal. It is at a level of half a wavelength for narrow (few \%) signal bands,
decreasing to $\approx 0.2 \lambda$ only for the broadest bandwidths expected
to be useful. This is a factor 10 less stringent than the rough value of
$\lambda/50$ used in the initial exploratory studies mentioned earlier, but exploiting this relaxed value will require post-processing of the
interferometer data in a computer.

The paper continues with a brief discussion on translating the two parameters 
identified into more familiar terms such as the range and bearing from one 
spacecraft to another. This discussion is sketchy and given only by
way of illustration, since the details of specific constellations are not
the main topic here. Finally, some implications of these results for the 
future design of free-flyer imaging systems are discussed. Conventional
``direct'' imaging systems, which form images directly on a panoramic detector 
(e.g.\ a CCD) at the focus of a (``dilute'') aperture, do not easily permit the 
use of the relaxed requirements on station-keeping described here. ``Indirect'' 
imaging systems, where the images are formed by post-processing of 
interferometer data in a computer, provide opportunities to make use of 
this (and other ancillary) information, and it is likely that future
high-resolution space-based imaging systems for astronomy will be of
the indirect type.

Background material on the topics of this paper can be found in several
excellent tutorials on the application of interferometry techniques to optical
astronomy. For example, see the volume edited by \citet{law00}\footnote{Lawson
also maintains the \textit{Optical Long Baseline Interferometry News} web site 
with a wealth of useful links and reference information at: \\
http://olbin.jpl.nasa.gov/papers/index.html}. Among others,
reviews of progress
and recent advances have been published by \citet{sha92} and by \citet{qui01}.

\section{The Imaging Process}

\begin{figure*}[ht!]
\epsscale{1.5}
\plotone{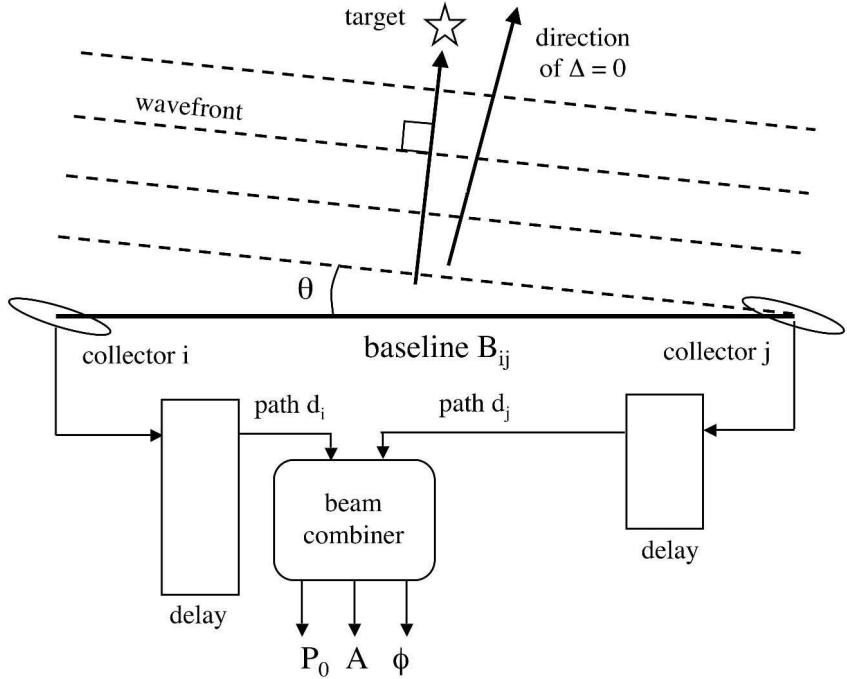}
\caption{The basic geometry of an interferometer. Two collectors $i$ and $j$
sample the wavefront from a very distant target at two points separated by
$B_{ij}^T = B_{ij} \cos \theta$ meters. The two samples are correlated
in the beam
combiner after experiencing relative path delays of
$B_{ij} \sin \theta + d_{i}$ and $d_{j}$, respectively. The beam combiner
calculates the fringe parameters, e.g.\ by scanning $\delta = d_{j} - d_{i}$
over one or two fringe periods, sampling, and fitting the pattern.
\label{fig:interf-sketch}}
\end{figure*}

How do we make images using data collected from astronomical targets with a
constellation of free-flying collectors? The relevant physics is described
mathematically in terms of the \textit{mutual coherence function} of the
wavefront, a quantity which can be measured with an interferometer, and the
\textit{Van Cittert-Zernike theorem}, which relates this mutual coherence 
function to the object on the sky (i.e.\ the target brightness distribution)
by a Fourier transform. There are numerous textbooks on this topic, see e.g.\ 
\cite{bw75, hec02}. The result relevant for now is that, whether
the imaging system is ``direct'', where an image is instantaneously formed on a 
2-dimensional detector (e.g.\ a dilute-aperture imager), or ``indirect'', where 
an image is formed by post-processing in a computer (e.g.\ an array of Michelson 
interferometers), the image is built up by summing \textit{fringe patterns} 
formed by the interference of light gathered from all pairs (\textit{i,j}) of 
collector elements in the constellation. One such pair, connected as a
Michelson interferometer, is sketched in Figure \ref{fig:interf-sketch}.

At any given wavelength $\lambda$, the interferometer fringe pattern has the
form (e.g.\ \citet{bw75}, Chapter 7):
\begin{equation} \label{eqn:narrowband-fringe}
P_{ij} = P_{0} [ 1 + A \cos(2 \pi k \Delta)],
\end{equation}
\noindent where $P_{0}$ is the total signal (photons, Joules, etc.)
collected by  the two elements, $k = 1/\lambda$, $A$ is the
\textit{fringe amplitude}, ($0 \leq A \leq 1$),
%
%
and $\Delta$ 
is the \textit{optical path difference} (often designated as OPD) composed of:
\begin{equation} \label{eqn:OPD}
\Delta = B_{ij} \sin\theta + d_{i} - d_{j}.
\end{equation}
\noindent $\Delta$ is the difference in total path (meters) from the wavefront
plane through the two ``arms'' to the beam combiner, consisting of
$B_{ij} \sin \theta + d_{i}$ on the left side of Figure
\ref{fig:interf-sketch}, and $d_{j}$ on the right. The angle $\theta$ (in 
radians, often assumed small so that $\sin \theta \approx \theta$) is measured
in the plane containing the interferometer baseline\footnote{This direction is
in fact on the surface of a cone with half-opening angle of
$(\pi/2) - \theta$ and axis coincident with the 
baseline $B_{ij}$. The ``ideal'' interferometer has a constant response on the 
surface of this cone, and the field of view must be further restricted
e.g.\ by adding an aperture stop into the optical system.}. As we shall see 
momentarily when we consider detection systems with finite bandwidth, it is 
important to keep $\Delta \approx 0$, so if $\theta \approx 0$ we shall need to 
have $d_{i} \approx d_{j}$. This can be achieved by bouncing the light around, 
in space or in a delay line, before it enters the beam combiner. For instance, 
if the constellation is deployed on a parabolic surface and the beam combiner is
located at the focus of this parabola, then the delays for each collector are 
``automatically'' adjusted to be (nearly) equal. This particular geometry is 
called a ``dilute aperture'' since it resembles a conventional reflecting
optical telescope with most of the collecting area removed. But this particular 
geometry is a special case and is not a requirement; the only requirement is 
that the optical paths somehow be kept nearly equal. We also define the 
\textit{fringe period} measured on the sky as $\lambda/B_{ij}^{T}$ radians.
Note that the fringe period \textit{increases} as the projected baseline
decreases with increasing $\theta$. Note also that a typical system at optical 
wavelengths may have $B_{ij}^{T} / \lambda \approx 1 \times 10^{9}$ and
$\theta \approx 3 - 100 \times 10^{-10}$ rads, or $\approx 60 - 2000\: \mu$as.

\begin{figure*}[ht!]
\epsscale{1.5}
\plotone{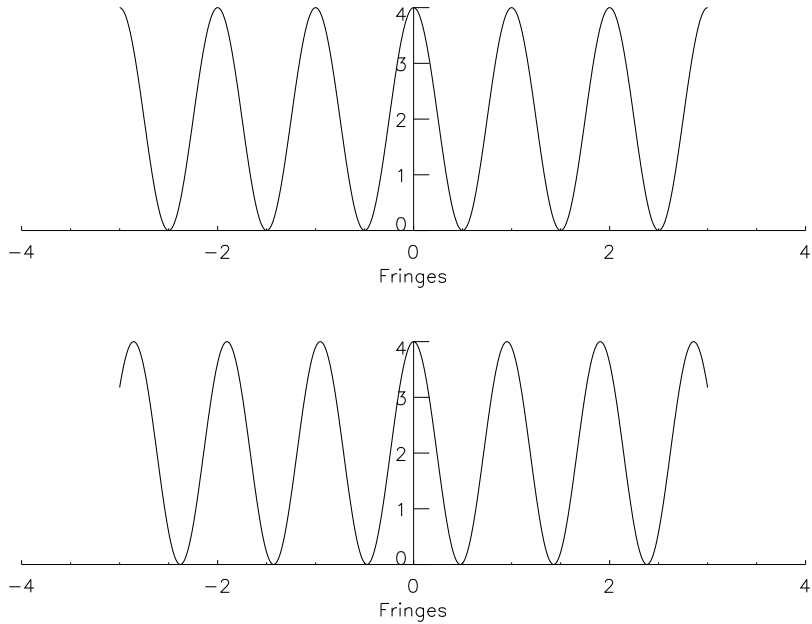}
\caption{\textit{Top panel:} Fringe pattern $P_{ij}$ from
Equation \ref{eqn:narrowband-fringe-2} for baseline $B$ as a function of
fringe phase $\phi_T$ in ``turns'' (see text). The instrument phase is assumed
to be at $\phi_I = 0$. \textit{Bottom panel:} Fringe pattern for a baseline
which is in reality 5\% larger. Notice that the error grows with angular
distance from the point where $\phi_I = 0$, reaching $\approx 0.15$ fringes
($54^{\circ} \approx 1$ rad) at the edge of this small 3-fringe field. The
y-axis is in units of the signal strength obtained by a single collector.
\label{fig:transverse}}
\end{figure*}

It is convenient to simplify the way we write the fringe pattern by defining
the quantities $\phi_T = k B_{ij} \sin \theta$ and $\phi_I = k (d_j - d_i)$,
both in units of ``turns'', so that Equation \ref{eqn:narrowband-fringe} becomes:
\begin{equation} \label{eqn:narrowband-fringe-2}
P_{ij} = P_{0} \{ 1 + A \cos[2 \pi ( \phi_T - \phi_I )]\},
\end{equation}
where $\phi_T$ is called the \textit{fringe phase} and depends on the location
of the target on the sky, and $\phi_I$ is the \textit{instrument phase},
which can be controlled by altering the internal delay difference $d_j - d_i$.
The elemental narrow-band fringe pattern of
Equation \ref{eqn:narrowband-fringe-2} is sketched as a function of instrument
phase in the top panel of Figure \ref{fig:transverse}. Note that the structure
of the fringes can be determined by scanning the instrument phase $\phi_I$ over
one or two turns and fitting the resultant fringe pattern to
Equation \ref{eqn:narrowband-fringe-2}, thereby obtaining the total power, the
fringe amplitude, and the fringe phase. This form of the equation also makes it
clear that errors in the instrument phase $\phi_I$, such as those arising from
station-keeping errors, directly masquerade as changes to the measured value
of fringe phase $\phi_T$, and therefore to errors in the measured position
$\theta$ of the target. Finally, Figure \ref{fig:transverse}
also makes it clear that the position of the target in the field of view
is obtained by \textit{counting fringes} (and fractions thereof) from a
reference point, for example the direction where $\phi_I = 0$.

The discussion so far has assumed that the interferometer is sensitive to only
one wavelength, i.e.\ that the bandwidth of the detection system is infinitely
narrow. The fringe pattern for a finite bandwidth detection system is the
weighted (vector) sum of the fringe patterns at all constituent wavelengths,
with amplitudes specified by the product of the source spectrum and the 
instrument response at each wavelength\footnote{This expression can be written 
as a Fourier Transform.}. A simple but illustrative case which is easy to 
calculate is for a rectangular bandpass of central wavelength $\lambda_{0}$ and 
width $\delta \lambda$, and a target spectrum that is flat over this wavelength 
range. In this case the resulting fringe pattern is:
\begin{equation} \label{eqn:broadband-fringe}
P_{ij} = P_{0} [ 1 + \frac{\sin(\pi \Delta / L_{c})}{\pi \Delta
/ L_{c}} \times A \cos(2 \pi k_{0} \Delta) ],
\end{equation}
\noindent where the symbols not already defined
in the previous two equations are
$k_{0} = 1/\lambda_{0}$, and $L_{c} = \lambda_{0}^{2}/\delta \lambda$. This
latter quantity is called the \textit{coherence length} for the light beam, a 
concept which arises as follows (\cite{bw75}, Ch.\ VII; \cite{hec02}, Ch.\ 7):
For a finite detector bandwidth of
$\delta \nu$ hertz, the signal in free space is a compact packet of many 
interfering waves with a coherence time of $\tau = 1 / \delta \nu$ seconds. 
Expressing the bandwidth in terms of wavelength, we have
$\delta \nu = -(c/\lambda_{0}^{2}) \times \delta \lambda$, so the coherence
time $\tau = (-1/c) \times (\lambda_{0}^{2} / \delta \lambda)$,
and $c \, \tau$ is a scale size of the wave packet called the coherence length,
$L_{c} = c \, \tau = \lambda_{0}^{2} / \delta \lambda$. The lower panel of
Figure \ref{fig:bandwidth} shows a sketch of this situation for
$\delta \lambda / \lambda_{0} = 0.2$. Note that, within a few fringes of the
central peak, the finite-bandwidth fringe in
the lower panel shows the same periodicity (and positioning, or fringe phase) as 
the single-wavelength case in the upper panel. This means that the relative 
placement of different components of the reconstructed target brightness 
distribution will be correct, but their amplitudes may be wrong. Perhaps even 
more serious, the signal-to-noise ratio of the amplitude measurement will be 
worse.

We are now in a position to begin the discussion on the accuracy with which the
relative positions of the collectors need to be known. For completeness, note
that we are here focussing only on that part of the error budget which arises
because of errors in the station-keeping. A full discussion of all error sources 
and their various contributions is, of course, dependent on the design details 
of each specific constellation.

\section{Two Requirements}

Inspection of Equations \ref{eqn:narrowband-fringe}, \ref{eqn:OPD}, and
\ref{eqn:broadband-fringe} above makes it clear that, in order to apply these 
equations to real data, there are two parameters which depend on station-keeping 
and which are important for each interferometer in the constellation:
the \textit {length of the projected baseline $B_{ij}^{T}$}, and the
\textit{optical path difference} $\Delta$ at the beam combiner. The accuracy
requirements on these  two parameters are presented below first by stating the
result, then providing  the derivation, and finally discussing some of the
implications.

\subsection{Baseline}
\label{subsec:baseline}

\textit{The fractional accuracy $\delta B_{ij}^{T} / B_{ij}^{T}$ with which the
projected distance between any i,j collector pair in the transverse plane must
be known is:}
\begin{equation} \label{eqn:baseline-error}
\delta B / B \lesssim 1 / (N_{f} \, Q_{T}).
\end{equation}
\noindent where $N_{f}$ is the number of fringes in the angle $\rho$,
$\rho$ is the angular radius of the field of view containing the object being
imaged, and $Q_{T}$ is the ``transverse quality factor'' expressing the
accuracy $\lambda/Q_{T}$ with which the ``optical figure'' of the imaging 
system is to be held. The super- and subscripts of $B$ have been dropped for 
convenience.

\subsubsection{Demonstration:}

The fringe pattern for an elemental interferometer expands and contracts about
the field center as B decreases or increases owing to station-keeping errors, 
and for a given (but in reality unknown) baseline error of $\delta B$ the 
resulting fringe phase error grows with angular distance from the center of the 
field. The situation for the narrow-band case is sketched in Figure
\ref{fig:transverse}.

Without loss of generality we can take the center of the fringe pattern to be
the centroid of the target brightness distribution at the operating wavelength
$\lambda$. Let $\rho$ be the angular radius (in radians) of a circle on the sky
which is just large enough to include all the emission from the target. The 
number $N_{f}$ of fringes in $\rho$ is $N_{f} = \rho/(\lambda/B)$. This number 
changes with small changes in the baseline length according to
$\delta N_{f} = \rho \, \delta B / \lambda$. Substituting for $\rho$ from the
initial expression for $N_{f}$ results in
$\delta N_{f} = N_{f} \times \delta B / B$. If we require that the optical
quality of the signal provided by this interferometer be better than a fraction 
$1/Q_{T}$ of a wavelength, then we must have $\delta N_{f} \lesssim 1/Q_{T}$. 
From this we derive $\delta B / B \lesssim 1 / (N_{f} \, Q_{T})$, which proves 
the result.

\subsubsection{Discussion:}

As an example, suppose the source is a distant star and the instrument is
intended to provide 30 independent resolution elements across the stellar 
diameter. Each resolution element is of angular size $\lambda / B_{max}$
radians, where $B_{max}$ is the longest (projected) baseline
in the constellation. 
Suppose further that it can be safely assumed there is no signal outside a
larger circle of some given angular radius, say
$\rho = 20 \times \lambda / B_{max}$ radians on the sky measured from the
center of the star.  For this example, $N_{f} \approx 20$. If we demand an 
``optical quality'' of, say, $\lambda/50$ in our imaging instrument, then
$Q_{T} = 50$ and $\delta B / B \lesssim 1/1000$. For a maximum baseline of
$B = 500$ meters, the largest \textit{unknown} $\delta B$ we can tolerate is
$\approx 0.5$ m. Note that since $N_{f} \propto B$ for a given source
size and operating wavelength, $\delta B $ is actually independent of $B$,
so \textit{the required accuracy in this example is the same for every baseline
in the constellation}. However, the stringency of the requirement grows with the 
size of the target field of view in the sky. For instance, for a supergiant star 
the desired FOV could be a factor of 10 larger than the example calculated 
above, and the required tolerance on any baseline would be $\approx 5$ cm. But 
now such a constellation would provide 300 independent resolution elements 
across the star. If the science goals permit, it may be more prudent to 
``shrink'' the overall extent of the constellation by a factor 10
in order to return to the scale of the 
initial example and relax the station-keeping requirements accordingly.

It is interesting to ask what the limit on $\delta B / B$ might be for very
large fields of view. For instance, for a Michelson radio interferometer with
a single ``feed horn'' at the focus of each collector, the
largest field of view will be limited by the diffraction pattern of the
individual collector elements. This field has an angular radius
$\approx \lambda / a$ where $a$ is the diameter of a collector element. In
this extreme case, $N_{f} \approx B/a$, so the maximum tolerable error is:
\begin{equation} \label{eqn:maxerror}
\delta B / B \lesssim 1 / (N_{f} \, Q_{T}) \approx a / (B \, Q_{T}) ;
\end{equation}
\noindent i.e., $\delta B \lesssim a / Q_{T}$, which depends only on the size
of the elements in the constellation. If a typical collector is one meter in
diameter and $Q_{T} = 50$, the required knowledge accuracy is $\approx 2$ cm.
The situation gets more demanding if a wide field
beam combiner is used, such as an optical Michelson interferometer with
detector pixels that are larger than the collector diffraction pattern,
or with many independent pixels in the detection system\footnote{Dilute
aperture ``Fizeau'' designs are not as optimum in signal-to-noise as are
the Michelson designs, and will not be discussed here any further.}.

To conclude this section, note that changes in operating wavelength have
similar effects on the array reponse to those discussed above and can
be modelled in the same way. One particularly deleterious effect of this is
that the point spread function of a system with a finite spectral resolution
will vary over the field of view, becoming radially elongated at large angular
distances. Such effects were first analysed in detail by radio astronomers
modeling the reponse of ground-based radio synthesis imaging arrays (see e.g.\
\citet{tho94}).

\subsection{Optical Path Difference}
\label{subsec:OPD}

\textit{The accuracy $\delta \Delta$ with which the optical path difference
must be controlled to zero is:}
\begin{equation} \label{eqn:OPD-error}
\delta \Delta / \lambda_{0} \lesssim (1 / Q_{OPD}) \times (\lambda_{0} / \delta \lambda).
\end{equation}
\noindent where $\delta \lambda / \lambda_{0}$ is the fractional bandwidth of
the signal and $Q_{OPD}$ is the ``OPD quality factor'', by analogy to the
$Q_{T}$ factor defined earlier. This requirement can also be written as
$\delta \Delta \lesssim L_{c} / Q_{OPD}$, where $L_{c}$ is the coherence length
defined following Equation \ref{eqn:broadband-fringe}.

\subsubsection{Demonstration:}

\begin{figure*}[ht!]
\epsscale{1.5}
\plotone{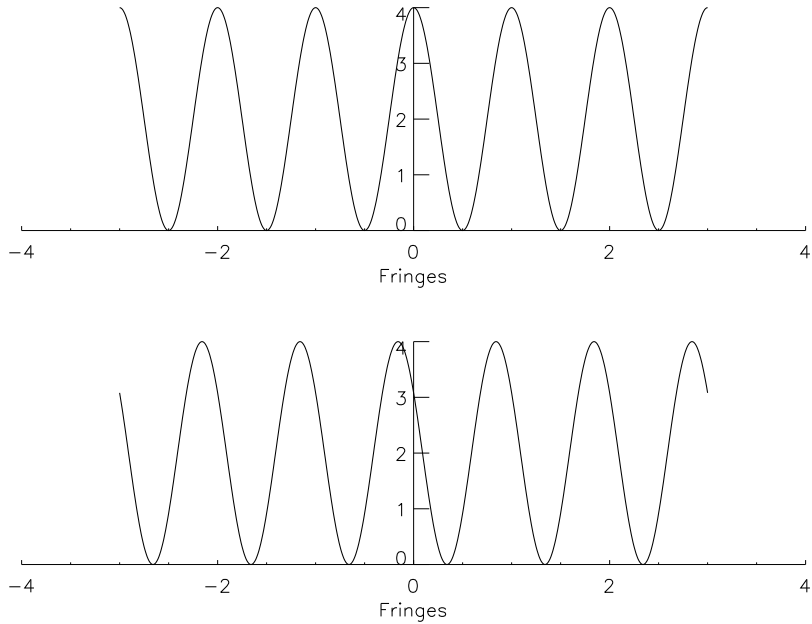}
\caption{\textit{Top panel:} Narrow-band fringe pattern from Figure
\ref{fig:transverse}.
\textit{Bottom panel:} Narrow-band fringe pattern shifted by a phase error of
$\delta \phi = 1$ rad, caused e.g.\ by an OPD error of
$\delta \Delta \approx \lambda / 6$. \label{fig:longitudinal}}
\end{figure*}

Some clarification is needed first, since the real problem here may not be
entirely obvious. It is clear from Equation \ref{eqn:narrowband-fringe} that
errors of $\delta \Delta / \lambda$ directly translate to a fringe error of
$G \, \delta \Delta / \lambda$ where $G$ is a geometry-dependent factor of
order unity depending on the specific geometry of the constellation\footnote
{For example, $G \approx 2$ for a classical (Fizeau) imaging system.}.
However, in this case it is not an expansion or contraction of the fringe
pattern about the 
field center, but a gross translation of the entire fringe pattern in a 
direction parallel to that of the baseline projection on the field of view. At 
first sight this is a serious problem, since it appears to require holding the 
collector positions to a tiny fraction of the wavelength. However, since
\textit{it is the whole fringe pattern that moved} and not the relative
separation between two points in the field of view, the angular distance
\textit{measured in fringes} between the center and the edge of the field
remains the same. In other words, the \textit{relative} fringe phase is
actually not sensitive to this shift! Figure \ref{fig:longitudinal} shows
a sketch of the situation.

\begin{figure*}[ht!]
\epsscale{1.5}
\plotone{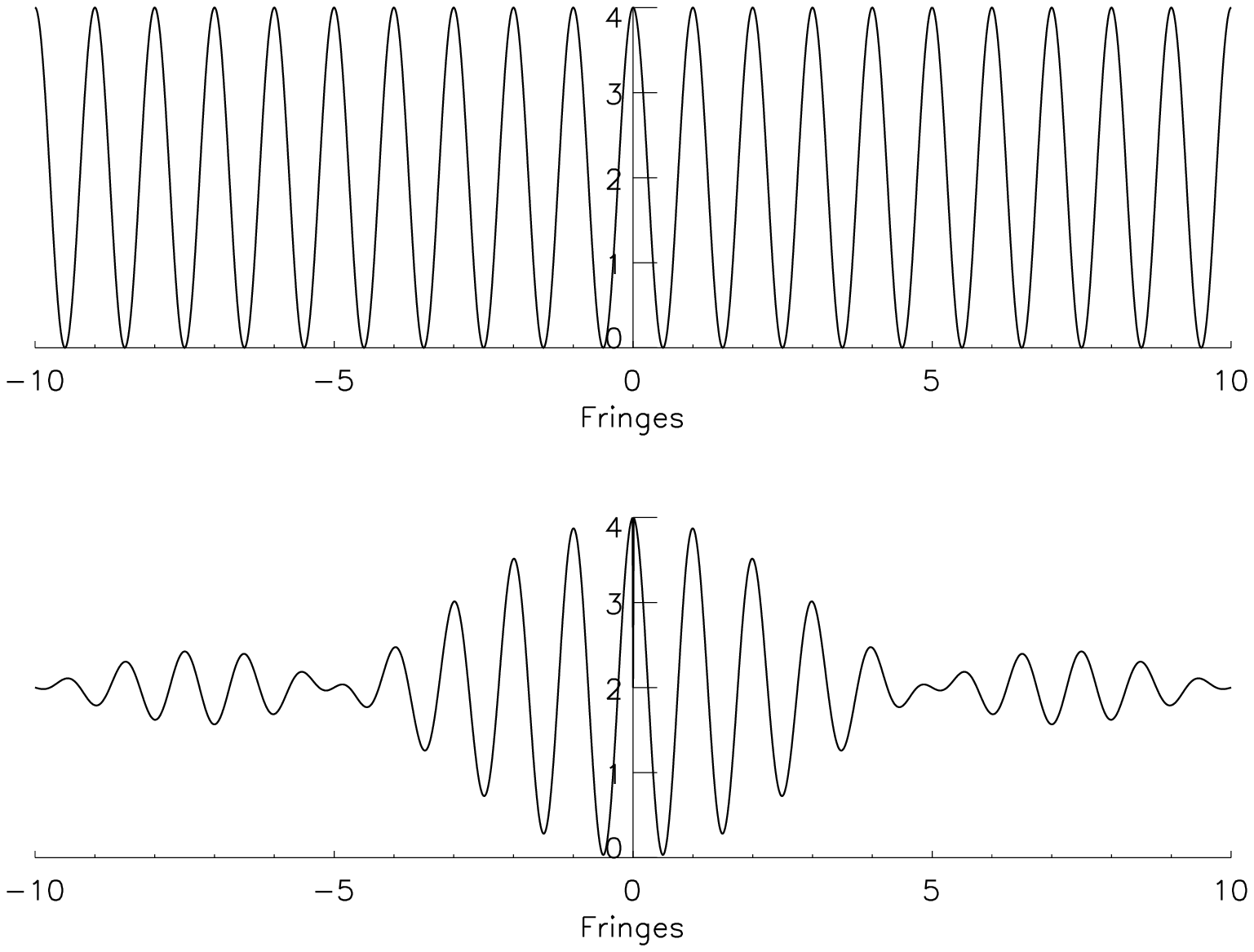}
\caption{\textbf{Top panel:} Fringe pattern as a function of OPD error
(in wavelengths) for a very narrow-band signal. \textbf{Bottom panel:}
The same fringe pattern
for a square bandpass of 20\%. Note how the fringes disappear beyond
$N_{f} \approx 5$. \label{fig:bandwidth}}
\end{figure*}

If we can design our system and/or (post-) process the data in order to use
this ``relative'' phase insead of an ``absolute'' phase, we would be entirely insensitive to OPD errors, although we would lose the ability to measure the 
target position in an absolute sense. Except for accurate wide-angle 
astrometry, no other applications in astrophysics require such absolute
position information. The real problem comes about because of a loss of fringe 
amplitude at large non-zero values of the OPD owing to the finite bandwidth of
the detection 
system. Only if the bandwidth is infinitely narrow will the fringe pattern 
amplitude remain independent of the OPD error $\delta \Delta$ as shown
e.g.\ in the upper panel of Figure \ref{fig:bandwidth}; the lower panel of
this figure shows a ``real'' fringe pattern for a finite (20\%) bandwidth 
following Equation \ref{eqn:broadband-fringe}. It is clear that, as the OPD 
error grows, so does the error in the measurement of the fringe amplitude. The 
scale size of the fringe pattern envelope is characterized by the coherence 
length $L_{c}$ as shown in Equation \ref{eqn:broadband-fringe}. If we can keep 
the OPD error $\delta \Delta$ to some fraction $1/Q_{OPD}$ of the coherence 
length $L_{c}$, then this amplitude error can be minimized and the measurement 
of fringe amplitude and fringe phase can be made subject only to photon noise
in the time scale of the drift in the station position. This requirement 
becomes $\delta \Delta \lesssim (1/Q_{OPD}) \times L_{c}$ or
$\delta \Delta / \lambda_{0} \lesssim (1/Q_{OPD}) \times (\lambda_{0} / \delta \lambda)$, which proves the result.

\subsubsection{Discussion:}

As an example, for $Q_{OPD} = 50$ and a bandwidth of a few \%, say
$\lambda_{0} / \delta \lambda \approx 30$, the control requirement on
$\delta \Delta$ is $0.6 \, \lambda$. This requirement will still be
difficult to achieve at optical wavelengths, but now
there is a better chance that on-board laser ranging 
systems may eventually provide adequate position knowledge without the use of 
signal photons, thereby allowing long integrations on faint sources. Bandwidths 
in excess of $\approx 10\%$ in a single detector channel are not likely to be 
used since the spectral dependence of the fringe pattern is often an important 
diagnostic for the on-board fringe detection algorithms. This means that the 
worst case for the control accuracy of the OPD (which is 
essentially the station-keeping requirement, as we shall see below) is
$\approx 0.2 \, \lambda_{0}$, still a factor of 10 less stringent than the
early estimates of $0.02 \, \lambda_{0}$ mentioned in the introduction to
this paper.

Now that the two major physical quantities have been identified, and their
accuracy requirements estimated, we can consider how to apply these criteria to real cases of specific mission concepts.

\section{Applications to Specific Mission Concepts}

The previous section has identified the two physical parameters $B$ and
$\Delta$ which characterize the constituent fringe patterns that go into 
building an image from data obtained with a constellation of free-flying 
collectors. In order to take the next step of translating the general knowledge 
requirements on $\delta B$ and $\delta \Delta$ into nanometers and microradians 
for a specific mission concept, we need to distinguish two cases depending on 
the brightness of the targets to be observed.

\subsection{Bright Targets}

If enough photons will be available to operate a servo loop to keep the optical
path difference $\Delta \approx 0$ to within some fraction of the coherence 
length $L_{c}$, then the control requirement on $\delta \Delta$ need not 
translate into a station-keeping requirement, but rather into a requirement on 
the total stroke and accuracy of the on-board delay line. In this case we are 
left only with the knowledge requirement on the projected baseline length which 
depends on the size of the target field of view, as stated in Equation
\ref{eqn:baseline-error}. The requirement on the accuracy in the range from one
spacecraft to another is of order half a meter. This still permits imaging of 
the stellar surface with a relative resolution of about 30. An error in the 
bearing of the second spacecraft with respect to the first in the interferometer 
pair also appears as an error in the projected baseline, but this component will 
be less significant; from the geometry of the interferometer in Figure
\ref{fig:interf-sketch} it is easy to show that an error of $\delta \alpha$
in the bearing translates into a fractional projected baseline error of
$\delta B / B \approx (\delta \alpha)^{2} / 2$. For the example taken in
\S \ref{subsec:baseline}, $\delta B / B \lesssim 1/1000$, so the
requirement on the bearing is 0.1 rad $\approx 1^{\circ}$ independent of B.

\subsection{Faint Targets}

If we want to observe faint (or partially resolved) targets which do not
provide enough photons to properly operate a control loop on the optical
path difference, then Equation \ref{eqn:OPD-error} drives the station-keeping
requirement. Errors in 
station-keeping map into errors in $\Delta$ in a manner dependent on the
specific constellation geometry. For instance, for a design where the beam combiner lies between the two collectors (as in the TPF-I concept),
a station-keeping error of $\delta B$ in the baseline length
will produce a delay error of $\delta \Delta = \delta B$, and we have seen from 
the demonstration following Equation \ref{eqn:OPD-error} that such errors need 
to be kept smaller than some fraction $1/Q_{OPD}$ of the coherence length.
If laser ranging is available and capable of measuring the baseline
length to sufficient precision, then that information could be fed to an
on-board delay line in the target signal path to correct for baseline
length errors. This would turn the requirement on $\delta B$ from a
\textit{control} requirement
into a \textit{knowledge} requirement\footnote{However, the
required \textit{precision} on that knowledge remains
a fraction $1/Q_{OPD}$ of the target signal coherence length.}.

Bearing errors $\delta \alpha$ also directly translate into delay errors; for
the SI concept of Figure \ref{fig:SIa-sketch} this is
$\delta \Delta \approx B \delta \alpha$, so the knowledge of bearing of the
second spacecraft with respect to the first is
$\delta \alpha \lesssim L_{c}/(B Q_{OPD})$ which for sample values used in the 
discussion in \S\ref{subsec:OPD} becomes
$\delta \alpha \lesssim 0.6 \, \lambda/B$ rads or
$\approx 0.1$ milli-arcseconds for
$B = 500$ m and $\lambda = 500$ nm. This will be a difficult 
requirement to meet if the collectors in the constellation are all more-or-less 
in a plane. Placing an additional ``metrology'' spacecraft some distance from 
this plane (e.g.\ in the ``Hub'' spacecraft for the SI design of
Figure \ref{fig:SIa-sketch}) would
alleviate the requirement by converting the inter-collector bearing measurement 
for station-keeping errors in the direction to the target (the ``L'' direction 
in Figure \ref{fig:SIa-sketch}) to a range measurement (with the usual required 
accuracy of order $L_{c}/Q_{OPD}$) from a collector to the metrology
spacecraft\footnote{Note that the problem remains of rendering the
station-keeping \textit{inertial}, in the sense described in the Introduction
to this paper. The presence of a ``hub'' spacecraft may present an
opportunity to address that problem as well.}.

\section{Indirect Imaging}

In order to take advantage of the relaxed control requirement on the optical
path difference described in Equation \ref{eqn:OPD-error}, we must find a method
of turning the recorded fringe amplitudes and (slowly-drifting) phases into an
image in such a way as to be insensitive to shifts of the whole fringe pattern 
owing to small OPD errors. There are several related ``post-processing''
techniques
which could be used for this. For instance, if the field of view contains a 
bright star besides the target of interest, fringe phases could be 
``referenced'' to this star. Another approach (first used at radio wavelengths) 
is to reference phases to a bright spectral line feature.

One quite general approach to this ``unstable-phase'' imaging problem which has 
been used with success for $\approx 40$ years in the radio astronomy community 
uses the concept of ``closure phase''. Almost 50 years ago, \citet{jen58}
presented a technique for measuring relative fringe phase which used three
radio-linked collectors coupled as three interferometers operating at a 
wavelength of 2.4 meters over baselines up to $\approx 10$ km. Owing to a
variety of instrumental problems related to the amplifiers and local oscillator 
electronics available at the time, the fringe phase between any two collectors 
was unstable and could normally not be measured; the source structure 
information had to be derived from the (squared) fringe amplitudes alone
(a familiar situation in present-day ground-based optical interferometry).
Jennison showed that, if the three observed fringe phases were summed, the
resultant combined phase was insensitive to equipment instabilities. With this
approach, \cite{jen59} showed that the brightness
distribution of the radio source Cygnus A, which until then was only known to be
elongated, actually consisted of two separated sources of nearly equal 
brightness straddling a peculiar optical object tentatively identified at the 
time with two galaxies in collision. This was the first observation to reveal 
the double-lobed structure of a powerful radio galaxy.

Applications of this method to circumvent atmospheric phase instabilities in 
optical interferometry were described by \citet{jen61} and, apparently 
independently, by \citet{rog68}. The first use of the term ``closure phase''
seems to be in the paper by \citet{rog74} describing an application 
at radio wavelengths using very stable and accurate, but independent, reference
oscillators at the three stations in a so-called ``very-long-baseline''
interferometer array. Since that time, closure phase has been used extensively
at radio, IR, and optical wavelengths, and there are many papers describing the 
subject, its virtues, and its limitations\footnote{For newcomers, the lectures 
presented at the \textit{Michelson Summer School}
by John Monnier and David Buscher 
in 2001, and by Peter Tuthill in 2003 are good sources; see
http://msc.caltech.edu/school/2003 and links there.}. \citet{mon03}
gives a review of optical interferometry in astronomy which includes a
discussion of closure phase, and also points out the intimate relationship
between it and the \textit{complex bispectrum}, a quantity constructed
in processing optical speckle interferometry data\footnote{\citet{cor87}
specifically (and elegantly) addresses the relationship between speckle
masking and phase closure.}. For the type of
collector constellations discussed in this paper, the phase of the complex
bispectrum is precisely the closure phase, and by (vector) averaging the
complex bispectrum the closure phases can be obtained in principle for
arbitrarily faint targets. One important point to make here is that use of such techniques requires that the fringe signals from each pair of collectors
be separable one from another, such that the individual interferometer fringe
phases can be uniquely assigned to the appropriate pair of collectors\footnote
{Note there is no requirement here that the set of baselines available in the
constellation be ``non-redundant'', although that may be useful for other
reasons.}. This leads in turn to requirements on the design of the optics at
the location where the collector beams are brought together to interfere.

\section{Conclusions}

The requirements on station-keeping for constellations of free-flying
collectors coupled as (future) imaging arrays in space for astrophysics
applications have been discussed. The typical \textit{knowledge} accuracy
required on the projected baseline of the instrument depends on the angular
size of the targets of interest; it is generally at a level of half a meter
for typical stellar targets, becoming of order a few centimeters
only for the widest attainable fields of view. The typical \textit
{control} accuracy required on the optical path difference depends on the 
wavelength and the bandwidth of the signal, and is at a level of half a 
wavelength for narrow (few \%) signal bands, becoming $\approx 0.2 \, \lambda$
for the broadest bandwidths expected to be useful.
If the fringe detection system is able to use signal photons or 
photons from a nearby reference star, then the most stringent requirement on 
station-keeping is that on the baselines, as described in
\S\ref{subsec:baseline}. If observations on faint targets are 
required, and fringe tracking is ``blind'', then the most stringent requirement 
on station-keeping is that on the optical path difference described
in \S\ref{subsec:OPD}. If on-board laser metrology is available and can
provide knowledge of the internal OPD to a fraction of the target signal
coherence length, then active fringe tracking could compensate for
station-keeping errors, even for faint sources.

In any case, the requirements are less severe than assumed in the early
studies of the problem.
The significance of this result is that, at this relaxed level of accuracy, it
may be possible to provide the necessary knowledge of array geometry without
the use of signal photons, thereby allowing observations of arbitrarily-faint 
targets.

Such constellations of free-flyers will produce images using various
computer-based image reconstruction techniques which relax the
level of accuracy required on the fringe phase stability of each component
interferometer. ``Closure-phase'' imaging is one such technique which
has been very successfully applied in ground-based radio and optical
astronomy in order to surmount instabilities owing to equipment and to the 
atmosphere. This technique appears to be directly applicable to space imaging 
arrays, where station-keeping drifts play the same role as atmospheric and 
equipment instabilities. More detailed modeling of anticipated station-keeping
errors and their effects on synthetic images is needed. For instance, there is
at present little justification for the values ``$Q_{OPD} = Q_{T} \approx 50$''
used repeatedly in this paper. A study needs to be done of the degradation in
image quality as these parameters are diminished (and the OPD and baseline errors grow), as well as a comparison of the merits of various image
reconstruction algorithms including the use of closure phase and bispectrum
analysis.

\section*{Acknowledgments}

I am grateful for stimulating discussions with Ken Carpenter, Dave Mozurkewich,
and Rick Lyon of the Stellar Imager Vision Mission team, and with R.\ Sridharan
of the Space Telescope Science Institute. Jesse Leitner of the Goddard
Space Flight Center provided helpful suggestions on an earlier version of this
paper. The comments of the referee have helped to clarify the presentation.
This work has been supported by the Space Telescope Science Institute
and by grants from the National Aeronautics and Space Administration.


\begin{thebibliography}{dummytext}
\bibitem[Born \& Wolf(1975)]{bw75}
Born, M., \& Wolf, E.~1975, \textit{Principles of Optics} (Fifth Edition) (Pergamon Press)
\bibitem[Cornwell(1987)]{cor87}
Cornwell, T.J.~1987, \aap, 180, 269
\bibitem[ESA(1996)]{esa96}
ESA Report SCI(96)7 1996, \textit{Kilometric Baseline Space Interferometry}, (European Space Agency, Paris)
\bibitem[Hecht(2002)]{hec02}
Hecht, E.~2002, \textit{Optics} (4th edition) (Addison Wesley)
\bibitem[Jennison(1958)]{jen58}
Jennison, R.C.~1958, \mnras, 118, 284
\bibitem[Jennison \& Latham(1959)]{jen59}
Jennison, R.C., \& Latham, V.~1959, \mnras, 119, 174
\bibitem[Jennison(1961)]{jen61}
Jennison, R.C.~1961, Proc.\ Phys.\ Soc.\ 78, 596 
\bibitem[Jones(1991)]{jon91}
Jones, D.~1991, in \textit{Technologies for Optical Interferometry in Space},
ed. S.\ P.\ Synnott (JPL Pub. D-8541), 177-181.
\bibitem[Lawson(2000)]{law00}
Lawson, P.R.~2000, \textit{Principles of Long Baseline Stellar Interferometry}
(JPL Pub. 00-009\footnote{Also at http://olbin.jpl.nasa.gov/papers/index.html})
\bibitem[Monnier(2003)]{mon03}
Monnier, J.D.~2003, Reports on Progress in Physics, 66, 789
\bibitem[Quirrenbach(2000)]{qui00}
Quirrenbach, A.~2000, ``Phase Referencing,'' in \textit{Principles of Long
Baseline Stellar Interferometry}, ed.\ P.\ R. Lawson
(JPL Pub. 00-009), 143
\bibitem[Quirrenbach(2001)]{qui01}
Quirrenbach, A.~2001, \araa, 39, 353
\bibitem[Rogstad(1968)]{rog68}
Rogstad, D.H.~1968, Applied Optics, 7, 585
\bibitem[Rogers et.al.(1974)]{rog74}
Rogers, A.E.E, et al.~1974, \apj, 193, 293
\bibitem[Shao \& Colavita(1992)]{sha92}
Shao, M., \& Colavita, M.M.~1992, \araa, 30, 457
\bibitem[Thompson(1994)]{tho94}
Thompson, A.R. 1994, ``The Interferometer in Practice,'' in \textit{Synthesis
Imaging in Radio Astronomy}, eds R.A.\ Perley, F.R.\ Schwab, \& A.H.\ Bridle, ASP Conference Series, Vol.\ 6 (Astron. Soc. of the Pacific, San Francisco), 11
%
\end{thebibliography}
\end{document}